# Spin exchange dominated by charge fluctuations of the Wigner lattice in the newly synthesized chain cuprate Na$_5$Cu$_3$O$_6$

Naveed Zafar Ali[1], Jesko Sirker[2*], Jürgen Nuss[1], Peter Horsch[1], and Martin Jansen[1*]

[1]Max Planck Institute for Solid State Research,

Heisenbergstrasse 1, D-70569 Stuttgart, Germany

[2]Department of Physics and Research Center OPTIMAS,

University of Kaiserslautern, D-67663 Kaiserslautern, Germany

Na$_5$Cu$_3$O$_6$, a new member of one dimensional charge ordered chain cuprates, was synthesized via the azide/nitrate route by reacting NaN$_3$, NaNO$_3$ and CuO. According to single crystal X-ray analysis, one dimensional $^1_\infty$CuO$_2^{n-}$ chains built up from planar, edge-sharing CuO$_4$ squares are a dominant feature of the crystal structure. From the analysis of the Cu–O bond lengths we find that the system forms a Wigner lattice. The commensurate charge order allows to explicitly assign the valence states of either +2 or +3 to each copper atom resulting in a repetition according to **Cu$^{2+}$–Cu$^{3+}$–Cu$^{2+}$–Cu$^{2+}$–Cu$^{3+}$–Cu$^{2+}$**. Following the theoretical analysis of the previously synthesized compounds Na$_3$Cu$_2$O$_4$ and Na$_8$Cu$_5$O$_{10}$, the magnetic susceptibility was expected to show a large dimer gap. Surprisingly, this is not the case. To resolve this puzzle, we show that the magnetic couplings in this compound are strongly affected by excitations across the Wigner charge gap. By including these contributions, which are distinct from conventional superexchange in Mott-insulators, we obtain a quantitative satisfying theoretical description of the magnetic susceptibility data.

# I. Introduction

Multinary oxides constitute a remarkably versatile and prolific class of materials. They have continued to play a major role in the fields of high temperature superconductivity (HTSC) [1] and colossal magneto resistivity (CMR) [2], or, more recently, multiferroics and spintronics [3]. Although, during the past decades, much effort has gone into unraveling the phenomena of HTSC in cuprates [4] and of CMR in manganates, no fully consistent and conclusive microscopic explanation has become available yet. The theoretical difficulty is due to the high complexity of the problems resulting among others from strong electron correlations, and coupled charge, spin, orbital and lattice degrees of freedom in collective systems. Furthermore, virtually all oxide materials showing HTSC or CMR include severe structural disorder, even decay into multiphase systems (phase separation, stripe formation) [5-8], a fact that has impeded theoretical analyses commonly relying upon translational invariance, and blurred experimental observations by inhomogeneous signal broadening effects. Thus, it would be highly desirable to employ fully periodic and chemically well defined materials as model systems for studying charge, spin and orbital ordering, either coupled or independent.

With the "azide/nitrate route" we have developed a rather efficient approach for the solid state synthesis of intrinsically doped multinary transition metal oxides [9, 10]. As a particular strength of this procedure, the oxygen content, and thus the valence state of the transition metal, can be precisely fixed by the alkali azide/nitrate ratio weighed in. At various illustrative examples, the "azide/nitrate route" has been proven to be rather versatile in providing highly defined materials showing interesting structural and physical properties. Among these, in particular, a new family of quasi one-dimensional intrinsically doped sodium cuprates(II/III) [11-16] has been found to be excellently suited for investigating the wealth of physical properties related. This includes the unique phenomenon of separation of spin and charge excitations [17], which constitutes potential for applications in non-linear photoelectric devices [18]. Furthermore, the close structural and electronic relationship of these materials to the high-temperature superconductors, and the known instabilities of the HTSC towards low-dimensional phenomena forms a background, which provides a strong motivation for close scrutiny of these materials [19-21].

The sodium cuprates (II/III) realized thus far, $Na_3Cu_2O_4$ and $Na_8Cu_5O_{10}$ [11, 12], are intrinsically doped Mott insulators. In both compounds the one dimensional $^1_\infty CuO_2^{n-}$ spin chains based on edge-sharing $CuO_4$ units with Cu—O—Cu bonds close to 90° are the dominating structural units. In contrast to many two-dimensional cuprates, these materials do not become superconducting upon doping. Instead, a charge-ordered state develops, in which spin-bearing divalent copper ions and non-magnetic Zhang-Rice singlets (holes/$Cu^{3+}$) [22] alternate with specific periodicities matching the hole filling factors of 1/2 and 2/5, respectively [11]. The two orthogonal oxygen *p*-orbitals overlap with the *d*-orbitals of the copper ions, thereby strongly reducing the hopping integrals, and the corresponding superexchange becomes very weak. The reduced kinetic energy explains, among others, why these doped edge-sharing 1D cuprates [$Na_{1+x}CuO_2$] are insulators.

Interestingly, these linear cuprates represent unambiguous manifestations of Wigner lattices (WL) with the charge ordering pattern being determined by long-range Coulomb interactions and distinct from a $4k_F$ charge-density wave [23-25]. Thereby the doped edge-sharing chain compounds exceed the high-Tc cuprates in correlation strength and provide a one-dimensional test ground for the study of charge stripe formation. The term WL is used here in the sense of a *generalized* WL [24] where electrons localize and form a superlattice on top of the underlying Cu-O lattice structure. Because of the strong charge localization these systems are examples of one-dimensional spin-½ Heisenberg chains with long-range exchange interactions which depend on the charge ordering pattern and the distance between the spins. It is well known that magnetism of high-$T_C$ compounds is controlled by superexchange and the motion of the doped holes which leads to a spin liquid state. It is this subtle interplay which makes the theoretical description of the magnetic properties in this regime so difficult, and the problem is still not fully understood. The charge excitations of the doped chains here are gapped due to the Wigner charge order. The excitations across the Wigner gap also contribute to the magnetism and compete with the standard superexchange processes that stem from excitations across the Mott-Hubbard gap. While such interplay has been proposed recently by theory [26], it is here for the first time possible to show that these 'Wigner exchange' processes [27] are indeed of qualitative importance for the description of the magnetic properties of a real compound.

The Wigner charge ordered compounds appear as doped relatives of the undoped, multiferroic chain compounds such as LiCuVO$_4$ and LiCu$_2$O$_2$ [28-31]. Chain cuprates with edge-sharing geometry show a number of intriguing magnetic properties. Due to the almost 90° angle of Cu—O—Cu bonds, the hopping $t_1$ between nearest neighbor Cu sites results mainly from direct Cu—Cu exchange while the next-nearest neighbor hopping $t_2$ originates mainly from the Cu—O—O—Cu path. Therefore the magnitude of the effective next-nearest neighbor magnetic exchange interaction $J_2$ is expected to be similar or even larger than that of the nearest-neighbor exchange $J_1$. Furthermore, $J_1$ tends to be ferromagnetic [32] while $J_2$ is antiferromagnetic leading, because of frustration, at least locally to a non-collinear spin structure. This non-collinearity is believed to be at the heart of the recently observed multiferroic behavior [33, 34] of the undoped edge-sharing chain cuprates LiCuVO$_4$ and LiCu$_2$O$_2$ [28-31] and has triggered a number of theoretical studies of the one-dimensional Heisenberg model with nearest and next-nearest neighbor interactions [35-38]. The question, however, what realistic values for the exchange constants $J_1$ and $J_2$ are, is far from settled. For LiCuVO$_4$, for example, recent neutron scattering data have been interpreted in terms of two weakly coupled antiferromagnetic chains, i.e, $J_2 \gg |J_1|$ [39]. Others, however, have argued that both interactions are of comparable magnitude [40], a view which is also supported by optical data [41] and an analysis of the susceptibility data [38].

Here, we report on Na$_5$Cu$_3$O$_6$, a new member of the family of mixed valent chain cuprates with a hole filling factor of 1/3. Based on the effective one-dimensional spin ½ Heisenberg model used to analyze the magnetic properties of Na$_3$Cu$_2$O$_4$ and Na$_8$Cu$_5$O$_{10}$, the magnetic structure of Na$_5$Cu$_3$O$_6$ is expected to be particularly simple: the spin-bearing divalent copper ions on next-nearest lattice sites within the chain should form dimers which are only weakly coupled among each other. Surprisingly, the magnetic susceptibility measurements show that this picture of weakly coupled dimers is incorrect. We discuss the reasons for the unexpected magnetic properties of Na$_5$Cu$_3$O$_6$ and present a quantitative theoretical analysis of the susceptibility data.

Our paper is organized as follows: In **Sec. II** we discuss the synthesis, experimental methodology and procedures employed. In **Sec. III** the crystal structures as well as the results for the magnetic susceptibility and specific heat are discussed. In **Sec. IV** we develop the theoretical model to describe the magnetism and discuss the differences to the previously synthesized charge ordered chain cuprates. **The last section** is devoted to a discussion of our results and conclusions.

## II. Experimental Details

***A. Material synthesis and characterization.*** $Na_5Cu_3O_6$ ($Na_{1.667}CuO_2$) has been prepared along the "azide/nitrate route", as a single phase microcrystalline powder **[9, 10]**. The starting materials, $NaNO_3$ (Merck, 99.99%), $NaN_3$ (Sigma-Aldrich, 99.99%), and CuO (prepared by heating $Cu(C_2O_4)·1/2\, H_2O$ in a flow of dry oxygen at 593 K, for 20 h) were mixed in the ratio required according to Eq. (1), further milled in a planet ball mill, pressed in pellets under $10^5$ N, dried under vacuum ($10^{-3}$ mbar) at 423 K for 12 h, and placed under argon in a closed steel container **[10]** provided with a silver inlay. In a flow of dry argon the following temperature profile was applied: 298→533 K (100 K/h); 533→673 K (5 K/h); 673→923 K (600 K/h); 923→943 K (200 K/h) and subsequent annealing for 50 h at 943 K.

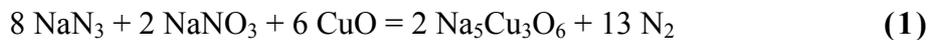

$$8\ NaN_3 + 2\ NaNO_3 + 6\ CuO = 2\ Na_5Cu_3O_6 + 13\ N_2 \qquad (1)$$

The temperature profile must strictly be followed to avoid any unpleasant circumstances. The obtained black powders, being very sensitive to humid air, were sealed in glass ampoules under argon atmosphere and all following manipulations with these substances were performed in inert atmospheres of purified argon. The X-ray investigation on powder samples was performed using a D8-Advance diffractometer (Bruker AXS, Karlsruhe, Germany) with Cu-K$\alpha_1$ radiation ($\lambda = 1.54178$ Å) at room temperature using a position-sensitive detector and a curved germanium monochromator.

Single crystals in the form of black needles can easily be singled out immediately after the reaction. However for better crystal quality the sample was post annealed at 873 K for 400 hours. Single crystal diffraction data were collected on a three circle diffractometer (Bruker AXS, Karlsruhe, Germany) equipped with a SMART-CCD (APEX I), at 293 K. The collection and reduction of data were carried out with the Bruker Suite software package **[42]**. Intensities were corrected for absorption effects applying a multi-scan method **[43]**.

The structure was solved by direct methods and refined by full matrix least-squares fitting with the SHELXTL software package [44]. Crystal structure data of $Na_5Cu_3O_6$ [45]: monoclinic, $P2_1/n$ (no. 14), $a = 5.706(2)$ Å, $b = 16.795(5)$ Å, $c = 8.113(3)$ Å, $\beta = 109.326(4)°$, $V = 733.6(4)$ Å$^3$, $Z = 4$, $\mu(MoK\alpha) = 8.896$ mm$^{-1}$, $\lambda = 0.71073$ Å, 11152 measured reflections, 3045 symmetry independent reflections ($2\theta_{max} = 69.68°$), 137 refined parameters, $R_1 = 0.065$, $wR_2 = 0.156$ (2402 $F_o > 4\sigma(F_o)$), $R_1 = 0.077$, $wR_2 = 0.163$ (all data). **Table 1** shows the atomic parameters and equivalent isotropic displacement parameters.

## B. Thermal analysis and magnetic measurements

The differential scanning calorimetry (DSC) measurements were carried out with a DSC device (DSC 404 C, Netzsch GmbH, Selb, Germany). The sample was heated at a rate of 30 K min$^{-1}$ in a corundum crucible under dry argon. The temperature dependence of the specific heat ($C_p$) of a polycrystalline sample of $Na_5Cu_3O_6$ was measured between 2 and 250 K using a commercial PPMS (Physical Property Measurement System, Quantum Design, 6325 Lusk Boulevard, San Diego, CA.) employing the relaxation method [46, 47]. To thermally fix the sample tablet (Ø = 5 mm and thickness 1mm) to the sapphire sample platform, a minute amount of Apiezon N vacuum grease was used. The heat capacity of the sample holder platform and grease was individually determined in a separate run and subtracted from the total measured heat capacities.

The magnetic susceptibility χ(T) is measured in the temperature range from 2 K to 680 K in magnetic fields up to 7 T using a SQUID-Magnetometer (MPMS 5.5, Quantum Design). For measurements above 350 K the sample was contained in warily dried SUPRASIL ampoule (Ø = 3 mm) that was long enough to extend over the coils of the magnetometer inside the oven.

In order to nullify the contribution, whatsoever, from spurious ferro- and paramagnetic impurities the Honda Owen correction [48, 49] was applied to the whole raw data obtained at 1, 3 and 7 Tesla, at magnetic field approaching infinity [H$^{-1}$ = 0]. The core electron diamagnetic susceptibilitiy has been calculated from the tabulated increment susceptibilities values [50, 51] whilst the van Vleck contributions which are positive and almost of the same order of magnitude as the diamagnetic contributions estimated from the energy differences of the orbitals and the spin-orbit coupling constant [52], both effects amounting about (~ -1.29 + 0.86) ×$10^{-4}$ emu/mol = — 0.43 ×$10^{-4}$ emu/mol, correspondingly.

## III. Experimental results

### A. Crystal Structure description

The crystal structure of $Na_5Cu_3O_6$ ($Na_{1.667}CuO_2$), a new member of the one-dimensional commensurate composite crystal family $Na_{1+x}CuO_2$ [11, 12], has been solved through single crystal X-ray diffractometry. Accordingly, the main structural characteristic is a one-dimensional polyanionic $^1_\infty CuO_2^{n-}$ chain, in which Cu is coordinated by oxygen in a square planar arrangement and these $CuO_4$-squares knit together in linear chains, sharing edges in trans-position, with mean intrachain Cu—Cu distances of 2.80 Å (**Fig. 1 (b)**).

Neighboring $CuO_{4/2}$ chains are stacked parallel to each other like in $MCuO_2$ cuprates formed by the bigger alkali metals ($M$ = K [53], Rb and Cs [54]). The linear chains in the title compound are shifted relative to each other by $b/2$, in contrast to the latter ones. The interchain Cu—Cu distances amount to 4.36 Å, on average. The sodium ions fill the space in-between the $CuO_{4/2}$ chains, in the form of layers of slightly elongated edge sharing $NaO_6$ polyhedra, with Na ions forming a tubular honeycomb-like arrangement (**Fig. 1 (a)**) with the channels occupied by cuprate ribbons.

The crystal structures of all members of chain cuprates belonging to the general family $Na_{1+x}CuO_2$ [11, 12], differ in the $Na/CuO_2$, and correspondingly in the $Cu^{2+}/Cu^{3+}$, ratios. These two features determine the periodicity/modulation along $b$ (chain direction). The resulting repetition unit for $Na_5Cu_3O_6$ is **$Cu^{2+}$–$Cu^{3+}$–$Cu^{2+}$–$Cu^{2+}$–$Cu^{3+}$–$Cu^{2+}$**. The $Cu^{3+}$ and $Cu^{2+}$ oxidation states can be clearly assigned according to the Cu—O bond distances, which are in the range of 1.855(4) - 1.881(3) for $Cu^{3+}$ and 1.896(3) - 1.936(3) Å for $Cu^{2+}$ (**Table 2**). The way of linking the primary structural units, as well as the variations of the copper to oxygen distances inevitably leads to deviations of the O—Cu—O angles from the ideal 90° and furthermore cause a slight undulation of the linear chains (∠Cu—Cu—Cu ≈ 177.5°) (**Table 3**). We note, that the alternation between $Cu^{3+}$ and $Cu^{2+}$ oxidation states is the hallmark of the generalized WL state, i.e., in contrast to a charge-density wave emerging from a Fermi surface instability [25].

In contrast to $NaCuO_2$ where all $Na^+$ ions are in the centers of the oxygen octahedra, in the title compound the sodium atoms are shifted off center, thereby giving freedom to accommodate more sodium atoms. This displacement in turn leads to two different oxygen environments for the Na atoms with Na–O bond length ranging from 2.270 to 2.767 Å.(**c.f Table 2**). The sodium atoms are shifted from the centers of the oxygen polyhedra in order to maximize the Na—Na distances. This leads in some cases to unusual thermal displacement parameters. This is true particularly for the position of Na5, which is thus better described applying a split position (Na5A and NA5B, **Table 1**).

Alternatively the structure of the title compound can also be interpreted within the 3+1D superspace approach [55,56,12], considering the structure as a composite one. $Na_5Cu_3O_6$ has the same small basic unit cell as $Na_3Cu_2O_4$ or $Na_8Cu_5O_{10}$ [$1/4 \times b(Na_3Cu_2O_4) \approx 1/5 \times b(Na_8Cu_5O_{10}) \approx 1/6 \times b(Na_5Cu_3O_6)$], the same superspace group, but a different modulation vector along the chain direction ($q = 5/6 \times b^*$).

## B. Thermal analysis and magnetic characterization

As monitored by differential scanning calorimetry there is a sharp reversible thermal signal at $T = 555$ K for $Na_5Cu_3O_6$, which can be assigned to the WL melting [25] (**Fig. 2**), in good accordance with the high temperature conductivity measurements [57]. The sample begins to decompose at about 1058 K, leaving mixtures of NaCuO [58], $NaCu_2O_2$ [59], and $Cu_2O$ [60] as the only solid residues. The phase purity of the sample was monitored by X-ray powder analysis as can be seen in **Fig. 3**.

The specific heat for $Na_5Cu_3O_6$ was recorded in the temperature range of 2 - 250 K. In the low temperature region one can see a $\lambda$-type anomaly at 23 K in the $C_p/T$(T) curve, as shown in **Fig. 4(a)**, which we assign to the onset of long-range AFM ordering. The ratio of $T(\chi_{max})/T(C_{max})$ is closer to the S = 1/2 Heisenberg model than to the Ising value [61]. Consequently, the absolute values of $C_p$ are uncertain and a substantial lattice contribution cannot be ruled out. To probe the nature of the specific heat anomaly at $T_N$ in more detail, we also displayed $C_p/T^2$ versus $T$ in **Fig. 4(a)** and we plotted the temperature derivative of the quantity $\chi_{mol} \times T$ ("Fisher's heat capacity", cf. ref. **[62]**) in **Fig. 4(b) [63, 64].** Both show a lucid picture with the Neel temperature precisely determined to be 22 K.

The magnetic susceptibility data is fitted by a Curie Weiss law (**c.f. Fig. 5**) in the temperature range of 150 to 680 K, giving a Curie constant of $C = 0.40$ emu K mol$^{-1}$ per Cu(II) and $\theta = -40.5$ K, corresponding to S = 1/2, which indicates a predominant antiferromagnetic interaction between Cu$^{2+}$ ions. $\mu_{eff}$ calculated from the Curie constant is 1.89 $\mu_B$ which is in good agreement with the spin only value of 1.73 $\mu_B$ expected for a Cu$^{2+}$ ($d^9$) system **[52]**. The susceptibility increases as temperature decreases down to ~30 K, where it has a rounded maximum. Below this temperature it shows a steep decrease with an inflexion point at $T_N = 23$ K which is, within the experimental error, in good agreement with the magnetic ordering transition temperature determined from the heat capacity measurements, $T_N = 23$ K.

## IV. Derivation of an effective spin model

### A. Long-range Coulomb interactions and charge order

In contrast to the corner-sharing geometry of copper-oxygen plaquettes as realized, for example, in the high-Tc cuprates, the edge-sharing geometry leads to strongly reduced hopping amplitudes. As a consequence, long-range Coulomb interactions within the chain

$$H_{Coul} = U \sum_j n_{j,\uparrow} n_{j,\downarrow} + \sum_j \sum_{d>0} V_d n_j n_{j+d} \qquad (2)$$

become important. Here $n_{j,\sigma}$ counts the number of electrons with spin $\sigma=\uparrow,\downarrow$ and $n_j = n_{j,\uparrow} + n_{j,\downarrow}$. Here U is the local Coulomb repulsion and $V_d=V/d$ represent the long-range Coulomb interactions defined in terms of the nearest-neighbor matrix element V and the distance d=1,2,3, … . The 1/d Coulomb form of the interaction is appropriate here, as we are dealing with insulating systems. The value of V is, however, screened by the static dielectric constant of the core electrons.

As the Coulomb repulsion exceeds the kinetic energy, the valence electrons form a Wigner crystal on top of the underlying Cu lattice and thereby minimize their Coulomb interaction.

For the hole doping concentration $x = 1/3$, as realized in $Na_5Cu_3O_6$, this leads to a charge localization with a unit cell (in a simplified picture considering only a single chain and only the copper atoms) comprising two $Cu^{2+}$ and one $Cu^{3+}$ ion as depicted in **Fig. 1(b).** The kinetic energy acts as a perturbation and the charge order is not frozen as the system will still undergo virtual charge excitations in order to take partial advantage of the kinetic energy,

$$H_{kin} = -\sum_{j,d,\sigma} t_d \left( c^+_{j,\sigma} c_{j+d,\sigma} + h.c. \right) \qquad (3)$$

where $c^+_{j,\sigma}$ and $c_{j,\sigma}$ are creation and annihilation operators for electrons at site j and spin σ, respectively, and the density operators are expressed as $n_{j,\sigma} = c^+_{j,\sigma} c_{j,\sigma}$. The resulting virtual transitions with hopping amplitudes $t_d$ lead to effective magnetic exchange interactions $J_d$ in a Heisenberg model with long-range interactions:

$$H_{HB} = \sum_j \sum_{d>0} J_d S_j S_{j+d} \qquad (4)$$

The positions of the spin operators $S_j$ representing the spin-1/2 of $Cu^{2+}$ are determined by the charge ordering pattern. The relevant exchange constants for $Na_5Cu_3O_6$ are shown schematically in **Fig. 6(a)**. Subsequently, in the effective spin model, Eq.(4), the sites with holes are unimportant and can be omitted, as seen in **Fig. 6(b,c).**

For $Na_3Cu_2O_4$ and $Na_8Cu_5O_{10}$ it was found that $J_2$ is antiferromagnetic and dominant **[25]**. Starting with the simplified magnetic structure as shown in **Fig. 6(b)** therefore suggests that the system consists of dimers which are coupled by weak ferromagnetic interactions. If this is the case, we can estimate $J_2$ by fitting the measured susceptibility at high temperatures to an independent dimer model. The result of an independent dimer fit is shown in **Fig. 5**. Whereby, we allow for a small constant contribution $\chi_0$ amounting to — 0.9 x $10^{-4}$ emu/mol, well within the range of what is expected due to the diamagnetic and the van Vleck temperature independent contributions to the magnetic susceptibility.

While the independent dimer model with $J_2$ = 145 K does fit the data well for high temperatures down to $T \sim 150$ K, it cannot describe the data at low temperatures. Here the theoretical model predicts an exponential decay of the susceptibility due to the large singlet-triplet gap of the dimer, $\Delta_D = J_2 = 145$ K, while the experimental data show a further increase of the susceptibility with decreasing temperature with a maximum at much lower temperatures T~30 K. The large excitation gap of the dimer model cannot be overcome by the other exchange couplings within the chain nor by the coupling between the chains J' which is expected to be of the order of the Neel temperature, J' ~ $T_N$ ~ 20 K. In this regard it is also important to note that the magnetic model shown in **Fig. 6(b)**, which does take the coupling of the dimers by a nearest-neighbor exchange $J_1$ into account, smoothly connects for increasing ferromagnetic coupling $J_1$ a dimer model (for $|J_1| \ll J_2$) with an effective antiferromagnetic S = 1 Heisenberg chain (for $|J_1| \gg J_2$). In the latter limit the model has a Haldane gap $\Delta_H$ ~ 0.4 $J_2$, i.e., in both limits the model shows a large gap and therefore cannot describe the experimental data.

The Curie fit shown in **Fig. 5**, on the other hand, clearly demonstrates that the dominant exchange interaction is antiferromagnetic. To understand this puzzle, we have to analyze the model $H = H_{Coul} + H_{kin}$ in detail. Naively, the exchange couplings $J_d$ are determined by charge fluctuations across the Hubbard gap, **Fig. 7(a),** involving a doubly occupied site in the virtual state and consequently $J_d \sim 4t_d^2/U$. Since $t_2$ is expected to be the dominant hopping amplitude due to the edge-sharing geometry, this approximation leads us to the dimer model shown in **Fig. 6(b)**.

## B. Virtual excitations across the Wigner charge gap

In contrast to a conventional Mott-Hubbard insulator, in a Wigner crystal there are also virtual excitations across the Wigner charge gap, as displayed in **Fig. 7(b),** which must be taken into account **[26]**. Careful analysis shows that these excitations also contribute to magnetism and can dramatically alter the exchange couplings $J_d$ as we will demonstrate in the following. A particular exchange process is shown in **Fig. 8(a).** The sequence of processes 1—2—3 leads to an interchange of the 2 electrons involved. Thus the process in **Fig. 8(a)** yields a contribution $J_2^D \sim 4t_1^2 t_2/D^2$, where D is the gap for *charge excitations across the Wigner gap*, which contributes to the 2$^{nd}$-neighbor magnetic coupling. The most remarkable features of these 'Wigner-type exchange' processes **[26,27]** are the following: (i) 'Wigner-type exchange' processes can get large as they do not have a U in the denominator, and it is possible that they overwhelm the superexchange process. (ii) In 'Wigner-type exchange' processes the hopping matrix elements may occur with odd powers, such that the character of the interaction depends on the signs of the involved hopping matrix elements. Thus 'Wigner-type exchange' can give rise to antiferromagnetic but also to ferromagnetic couplings. The latter comes unexpected for a transition metal oxide without orbital degeneracy and in absence of Hund coupling. We also note that the intermediate states in **Fig. 8(a)** after steps 1 and 2 are at different excitation energies $D_1$ and $D_2$, respectively. To simplify the presentation we shall adopt here an average Wigner gap D for a particular compound. We stress, however, that different compounds with different charge order naturally have different D. Starting from the ordered state, this energy scale can easily be calculated. Taking the first three non-vanishing Coulomb terms into account we find $D_{Na5} = V_2 - 2V_3 + V_4$ for $Na_5Cu_3O_6$. With $V_d = V/d$ this leads to $D_{Na5} = V/12$. For $Na_3Cu_2O_4$ with hole doping concentration $x = 1/2$, on the other hand, we find using the same approximation $D_{Na3} = V_1 - 2V_2 + 2V_3$ leading to $D_{Na3} = 2V/3$. The energy scale for charge fluctuations is therefore smaller for $Na_5Cu_3O_6$ than it is for $Na_3Cu_2O_4$. Since D enters quadratically in $J_2^D$, this has a dramatic effect. In this respect it is also important to note that there is a qualitative difference between the two compounds. While the charge order in $Na_3Cu_2O_4$ is stabilized by $V_1$, the next-nearest neighbor interaction $V_2$ is required for stability in the case of $Na_5Cu_3O_6$. In addition to the Coulomb interactions within the chain, also interchain Coulomb interactions contribute to the stability of the charge order and have to be taken into account when calculating the excitation energy D. This makes a precise determination of D difficult. As a rough estimate we find that $D_{Na5} = 0.2 - 0.6$ V while $D_{Na3} = 0.8 - 1.3$ V.

The various exchange processes in $Na_5Cu_3O_6$ lead to

$$J_1 = 4\frac{t_1^2}{U} + 8\frac{t_1^2 t_2}{D^2} + 16\frac{t_1^2 t_2}{UD} + 16\frac{t_1^2 t_2^2}{D^3}; \quad \quad 5(a)$$

$$J_2 = 4\frac{t_2^2}{U} + 4\frac{t_1^2 t_2}{D^2} + 8\frac{t_1^2 t_2}{UD}; \quad \quad 5(b)$$

$$J_3 = 4\frac{t_3^2}{U} + 4\frac{t_1 t_2 t_3}{D^2} + 8\frac{t_1 t_2 t_3}{UD} \quad \quad 5(c)$$

For $J_1$ we have also included the 4$^{th}$ order process involving charge fluctuations at two different $Cu^{3+}$ sites. In addition, we expect a ferromagnetic contribution to $J_1$ due to Hund's coupling at the oxygen sites [32]. We note that Eqs. 5(a-c) are obtained by an expansion in powers of $t_d/U$ and $t_d/D$. While this is still justified even for the relatively small Wigner gap in $Na_5Cu_3O_6$ with $t_d/D < ½$ (see also below), such an expansion will eventually break down for compounds with more complicated charge orders and thus smaller Wigner gaps. In such a case we will be again confronted with the full problem as in a correlated metal, Fig.**7(c)**, where the mobile charges interact strongly with the magnetic degrees of freedom, i.e., as in the high-Tc compounds. For the case of robust charge order considered here, we can use Eqs. 5(a-c) to estimate the superexchange constants $J_d$ with the role played by the exchange terms across the Wigner gap being determined by the *sign of the hopping amplitudes* $t_1$, $t_2$, and $t_3$.

From band structure calculations for the edge-sharing chain cuprates $Li_2CuO_2$ **[65]** and $LiCu_2O_2$ **[66]** it follows that the hopping amplitudes $t_1$, $t_2$ are both negative (note the convention in the definition of the kinetic energy (3) with the minus sign). The sign of $t_3$ has not been determined in these works but the phases of orbital overlaps suggest that $t_3>0$. Estimates for the parameters in Eq. (5) – except for the hopping amplitude $t_3$ - have been obtained by an analysis of optical data for $LiCuVO_4$ **[41]** leading to U ~ 3.75 eV, V~1.6 eV, $t_1$ ~ -0.08 eV, and $t_2$ ~ –0.1 eV. In **Fig. 9** we show the exchange constants as given in Eq. **(5)** as a function of D for U ~ 3.5 eV, $t_1$ ~ -0.08 eV, $t_2$ ~ –0.12 eV, and *assuming* $t_3$ ~ 0.05 eV.

**Fig. 9** clearly shows that virtual excitations across the Wigner gap, while negligible for $Na_3Cu_2O_4$, play an important role for $Na_5Cu_3O_6$ and lead to a much smaller antiferromagnetic or even small ferromagnetic coupling $J_2$. As example, we take D = 0.5 eV as reasonable for $Na_5Cu_3O_6$ with all other parameters as given in the caption of Fig. 9 and obtain $J_1$ ~ –145 K, $J_2$ ~ 10 K, $J_3$ ~ 150 K. This means that $J_3$ is the dominant antiferromagnetic interaction in sharp contrast to $Na_3Cu_2O_4$ where with D = 0.9 eV we obtain $J_2$ ~ 130 K in good agreement with the previous theoretical analysis in Ref. **[25]**. Note that in the latter case the exchange paths $J_1$ and $J_3$ do not exist due to the charge ordering pattern.

## *C. Numerical results for the magnetic susceptibility*

The exchange constants we have found for $Na_5Cu_3O_6$ imply that the magnetic structure of a single chain can be understood as a two-leg ladder with antiferromagnetic couplings along the legs and ferromagnetic coupling along the rungs as shown in **Fig. 6(c)**. Such a system will also show an excitation gap, however, in this case the gap is only of order $\Delta$ ~ 0.2 $|J_1|$ ~ 20 K **[67]** and can easily be overcome by interchain couplings.

To calculate the susceptibility for the Heisenberg model as depicted in **Fig. 6(c)**, we have used a density-matrix renormalization group algorithm applied to transfer matrices. This algorithm allows it to perform the thermodynamic limit exactly. The density-matrix renormalization group is used to extend the transfer matrices in imaginary time direction (corresponding to a successive lowering of the temperature) while keeping the number of states in a truncated Hilbert space fixed. For details the reader is referred to Refs. **[68-74]**. In **Fig. 10** we show a fit of the experimental data with $J_1$ = - 43 K, $J_2$ = 0 K, $J_3$ = 69 K. In addition, we allow for a small constant contribution $\chi_0$ amounting to — 1.6 x $10^{-4}$ emu/mol. This is not far off from the estimate obtained in Sec. IIB. We note that exchange constants of this magnitude are obtained from Eq. (5) if we choose, for example, U = 3.5 eV, $t_1$ = -0.08 eV, $t_2$ = –0.08 eV, $t_3$ = 0.04 eV, and D=0.6 eV and are therefore consistent with electronic structure calculations and the considerations above. Furthermore, we want to remind the reader that different Wigner gaps occur for different intermediate states and that the use of an average gap D is a stark simplification. The small deviations when comparing the fit and the experimental data in **Fig. 10** might be related to non-negligible interchain as well as longer-ranged intrachain couplings.

# V. Conclusion

In this article, the synthesis via the azide/nitrate route, structure determination from single-crystal data, as well as thermal and magnetic properties of the new member of mixed-valent sodium cuprates (II,III), $Na_5Cu_3O_6$ , are presented. From the structural analysis we find that this chain cuprate forms a commensurate Wigner lattice at temperatures T<555 K with a repetition pattern **$Cu^{2+}$–$Cu^{3+}$–$Cu^{2+}$–$Cu^{2+}$–$Cu^{3+}$–$Cu^{2+}$.** Unexpectedly, the experimental data for the susceptibility turn out to pose a serious challenge for our theoretical understanding of the magnetic exchange processes in charge ordered cuprates. While a Curie fit shows that the dominant exchange is antiferromagnetic, no spin excitation gap - expected if the next-nearest neighbor interaction is antiferromagnetic and dominant as in other charge ordered chain cuprates – has been found. To resolve this puzzle, we have argued that virtual excitations across the Wigner gap become important in this new compound with hole doping 1/3 while they have only a negligible effect in systems such as $Na_3Cu_2O_4$ where the hole doping is 1/2 and the charge order therefore more stable. Using realistic parameters for the Coulomb energies and hopping amplitudes we have presented a detailed analysis of the magnetic exchange constants in charge ordered chain cuprates as a function of the Wigner gap. This analysis has shown, in particular, that virtual excitations across the Wigner gap lead to a dominant antiferromagnetic coupling between third nearest neighbors for $Na_5Cu_3O_6$ while the next-nearest neighbor coupling is much smaller and possibly even ferromagnetic. A numerical calculation of the susceptibility for an effective long-range Heisenberg model with parameters as obtained from this analysis was finally shown to lead to a good agreement with the experimental data. We therefore conclude that $Na_5Cu_3O_6$ is the first charge ordered compound where the importance of virtual excitations across the Wigner gap has been convincingly demonstrated.

## Acknowledgments

Special thanks to Mrs. Eva Brücher for susceptibility measurements, and Mrs. Gisela Siegle for heat capacity measurement (MPI-FkF). J.S. acknowledges support by the graduate school of excellence MAINZ (MATCOR).

# References

1: T. Mayer, M. Eremin, I. Eremin, and P. F. Meier, J. Phys.: Condens. Matter **19**, 116209, (2007).

2: C. N. R. Rao, R. Mahesh, A. K. Raychaudhuri, and R. Mahendiran, J. Phys. Chem. Solids **59** (**4**), 487, (1998).

3: C. Felser, G. H. Fecher, and B. Balke, Angew. Chem. Int. Edn. **46**, 668, (2007).

4: H. Takagi, Nature Materials **6**, 179, (2007).

5: V.V. Moshchalkov, L. Trappeniers, and J. Vanacken, Physica C: Superconductivity **887**, 341-348, (Part 2), (2000).

6: J. M. Tranquada, B. J. Stemlieb, J. D. Axe, Y. Nakamura, and S. Uchida, Nature (London) **375**, 561 (1995).

7: N. Kumar, and C. N. R. Rao, ChemPhysChem. **4**(5), 439, (2003).

8: Colin V. Parker, Pegor Aynajian, Eduardo H. da Silva Neto, Aakash Pushp, Shimpei Ono, Jinsheng Wen, Zhijun Xu, Genda Gu & Ali Yazdani, Nature **468**, 677 (2010).

9: D. Trinschek, and M. Jansen, Angew. Chem. Int. Edn. **38**,133, (1999).

10: M. Sofin, E.-M Peters, and M. Jansen, Z. Anorg. Allg. Chem. **628**, 2697, (2002).

11: M. Sofin, E.-M. Peters, and M. Jansen, J. Solid State Chem. 178 (12), 3708, (2005).

12: J. Nuss, S. Pfeiffer, S. van Smaalen, and M. Jansen, Acta Crystallogr. B**66**, 27, (2010).

13: L. Capogna, M. Mayr, P. Horsch, M. Raichle, R. K. Kremer, M. Sofin, A. Maljuk, M. Jansen, and B. Keimer, Phys. Rev. B **71**, 140402(R) (2005).

14: M. Mayr, and P. Horsch, Phys. Rev. B **73**, 195103, (2006).

15: M. Raichle, M. Reehuis, G. Andre, L. Capogna, M. Sofin, M. Jansen, and B. Keimer, Phys. Rev. Lett. **101**, 047202 (2008).

16: L. Capogna, M. Reehuis, A. Maljuk, R. K. Kremer, B. Ouladdiaf, M. Jansen, and B. Keimer, Phys. Rev. B **82**, 014407 (2010).

17. M. Daghofer and P. Horsch, Phys. Rev. B **75**, 125116 (2007).

18. S. Maekawa, T. Tohyama, S.E. Barnes, S.Ishihara, W. Koshibae and G. Khalilullin, *Physics of Transition Metal Oxides* (Springer, Berlin , 2004) pp. 95 .

19: C. Kim, A. Y. Matsuura, Z. Shen, N. Motoyama, H. Eisaki, S. Uchida, T. Tohyama, and S. Maekawa, Phys. Rev. Lett. **77**(19) 4054, (1996).

20: M. Coey, Nature **430**, 155, (2004).

21: S. A. Kivelson, I. P. Bindloss, E. Fradkin, V. Oanesyan, J. M. Tranquada, A. Kapitulnik, and C. Howald, Rev. Mod. Physics **75**, 1201, (2003).

22: F. C. Zhang, and T. M. Rice, Phys. Rev. B **37**, R3759, (1988).

23: E. Wigner, Phys. Rev. **46**, 1002, (1934).

24: J. Hubbard, Phys. Rev. B **17**, 494 (1978).

25: P. Horsch, M. Sofin, M. Mayr, and M. Jansen, Phys. Rev. Lett. **94**, 076403, (2005).

26: M. Daghofer, R. M. Noack, and P. Horsch, Phys. Rev. B 78, 205115 (2008).

27: We decided to introduce here 'Wigner exchange' instead of 'kinetic exchange' as in Ref. **[26],** as the latter notion strictly speaking also includes conventional superexchange.

28: T. Masuda, A. Zheludev, A. Bush, M. Markina, and A. Vasiliev, Phys. Rev. Lett. **92**, 177201 (2004).

29: S. Park, Y. J. Choi, C. L. Zhang, and S.-W. Cheong, Phys. Rev. Lett. **98**, 057601 (2007).

30: S. Seki, Y. Yamasaki, M. Soda, M. Matsuura, K. Hirota, and Y. Tokura, Phys. Rev. Lett. **100**, 127201 (2008).

31: F. Schrettle, S. Krohns, P. Lunkenheimer, J. Hemberger, N. Büttgen, H.-A. Krug von Nidda, A. V. Prokofiev, and A. Loidl, Phys. Rev. B **77**, 144101 (2008).

32: The ferromagnetic contribution to $J_1$ arises from a $p^4$ intermediate state on O and the action of Hund coupling **[25]**.

33: H. Katsura, N. Nagaosa, and A. V. Balatsky, Phys. Rev. Lett. **95**, 057205 (2005).

34: I. A. Sergienko and E. Dagotto, Phys. Rev. B **73**, 094434 (2006).

35: T. Vekua, A. Honecker, H.-J. Mikeska, and F. Heidrich-Meisner, Phys. Rev. B **76**, 174420 (2007).

36: J. Sudan, A. Lüscher, and A. M. Läuchli, Phys. Rev. B **80**, 140402 (R) (2009).

37: F. Heidrich-Meisner, I. P. McCulloch, and A. K. Kolezhuk, Phys. Rev. B **80**, 144417 (2009).

38: J. Sirker, Phys. Rev. B **81**, 014419 (2010).

39: M. Enderle, B. Fåk, H.-J. Mikeska, R. K. Kremer, A. Prokofiev, and W. Assmus, Phys. Rev. Lett. **104**, 237207 (2010).

40: S.-L. Drechsler, S. Nishimoto, R. Kuzian, J. Malek, W. E. A. Lorenz, J. Richter, J. van den Brink, M. Schmitt, and H. Rosner, arXiv:1006.5070

41: Y. Matiks, P. Horsch, R. K. Kremer, B. Keimer, and A. V. Boris, Phys. Rev. Lett. **103**, 187401 (2009).

42: Bruker Suite, Version 2008/3, Bruker AXS Inc., Madison, USA, (2008).

43: G. M. Sheldrick, SADABS, Bruker AXS area detector scaling and absorption, Version 2008/1, University of Göttingen, Germany, (2008).

44: G. M. Sheldrick, Acta Crystallogr. A**64**, 112, (2008).

45: Further details may be obtained from Fachinformationszentrum Karlsruhe, 76344 Eggenstein-Leopoldshafen, Germany [Fax: (+49)-7247-808-666; E-Mail: crysdata(at)fiz-karlsruhe.de, http://www.fiz-karlsruhe.de/request for deposited data.html] on quoting the CSD number: 422413.

46: R. Bachmann, F. J. DiSalvo Jr., T. H. Geballe, R. L. Greene, R. E. Howard, C. N. King, H. C. Kirsch, K. N. Lee, R. E. Schwall, H. U. Thomas, and R. B. Zubeck, Rev. Sci. Instrum. **43**, 205 (1972).

47: J. S. Hwang, K. J. Lin, and C. Tien, Rev. Sci. Instrum. **68,** 94, (1997).

48: K. Honda, Ann. Phys. **32**, 1027, (1910).

49: M. Owen, Ann. Phys. **37**, 657, (1910).

50: Landolt-Börnstein, Zahlenwerte und Funktionen aus Naturwissenschaften und Technik; Neue Serie, Gr. II, Bd. 2, Springer, Berlin, (1966).

51: P. W. Selwood, Magnetochemistry, 2nd ed. Interscience, New York, p. 78, (1956).

52: H. Lueken, Magnetochemie, Teubner, Leipzig, (1999).

53: G. A. Costa, and E. Kaiser, Thermochimica Acta **269**, 591, (1995).

54: K. Hestermann, and R. Hoppe, Z. Anorg. Allg. Chem. **367**, 249, (1969).

55: S. van Smaalen, R. E. Dinnebier, M. Sofin, and M. Jansen, Acta Crystallogr. B**63**, 17, (2007).

56: S. van Smaalen, Z. Kristallogr. **219**, 681, (2004).

57: Naveed Zafar Ali, Ph.D. Dissertation, Max-Planck-Institut für Festkörperforschung, Stuttgart, (2011).

58: R. Hoppe, H. Hestermann, and F. Schenck, Z. Anorg. Allg. Chem. **367**, 275, (1969).

59: G. Tams, and Hk. Müller-Buschbaum, J. Alloy. Compd. **189**, 241, (1992).

60: A. Kirfel, and K.D.Eichhorn, Acta Crystallogr. A**46**, 271, (1990).

61: L. J. de Jongh and A. R. Miedema, Advances in Physics **50**(8), 947, (2001).

62: M. E. Fisher, Philos. Mag. **7**, 1731 (1962).

63: A. Möller, U. Löw, T. Taetz, M. Kriener, G. André, F. Damay, O. Heyer, M. Braden, and J. A. Mydosh, Phys. Rev. B **78**, 024420 (2008).

64: O. Janson, W. Schnelle, M. Schmidt, Yu. Prots, S.-L. Drechsler, S. K. Filatov and H. Rosner, New J. Phys., **11**, 113034 (2009).

65: R. Weht and W. E. Pickett, Phys. Rev. Lett. **81**, 2502 (1998).

66: A. A. Gippius et al., Phys. Rev. B **70**, 020406(R) (2004).

67: S. Larochelle, and M. Greven, Phys. Rev. B. **69**, 092408 (2004).

68: R. Bursill, T. Xiang, and G. Gehring, J. Phys. Condens. Matter **8**, L583 (1996).

69: X. Wang, and T. Xiang, Phys. Rev. B **56**, 5061 (1997).

70: N. Shibata, J. Phys. Soc. Jpn. **66**, 2221 (1997).

71: J. Sirker and A. Klümper, Phys. Rev. B **66**, 245102 (2002).

72: J. Sirker and A. Klümper, Europhys. Lett. **60**, 262 (2002).

73: J. Sirker, N. Laflorencie, S. Fujimoto, S. Eggert, and I. Affleck, Phys. Rev. Lett. **98**, 137205 (2007).

74: J. Sirker, N. Laflorencie, S. Fujimoto, S. Eggert, and I. Affleck, J. Stat. Mech. **P02015** (2008).

**Table 1.** Atomic coordinates and equivalent isotropic displacement parameters $U_{eq}$ (Å$^2$) for Na$_5$Cu$_3$O$_6$. $U_{eq}$ is defined as one third of the trace of the orthogonalized U$_{ij}$ tensor.

|      | x          | y           | z          | $U_{eq}$   |
|------|------------|-------------|------------|------------|
| Cu1  | 0.2366(1)  | 0.04412(3)  | 0.76471(7) | 0.0117(2)  |
| Cu2  | 0.2380(1)  | –0.12180(3) | 0.75447(7) | 0.0129(2)  |
| Cu3  | 0.2477(1)  | 0.21046(3)  | 0.75715(7) | 0.0119(2)  |
| Na1  | –0.2284(4) | 0.0466(1)   | 0.8780(3)  | 0.0185(4)  |
| Na2  | –0.2272(4) | –0.1331(1)  | 0.8807(3)  | 0.0155(4)  |
| Na3  | –0.2833(5) | –0.2695(2)  | 0.5997(4)  | 0.0427(9)  |
| Na4  | –0.2866(5) | –0.0537(3)  | 0.5591(3)  | 0.049(1)   |
| Na5A | 0.214(1)   | 0.3582(4)   | 1.0574(6)  | 0.0253(9)  |
| Na5B | 0.719(1)   | 0.1862(4)   | 0.6055(8)  | 0.030(1)   |
| O1   | 0.4223(7)  | –0.0384(2)  | 0.6959(5)  | 0.0164(6)  |
| O2   | 0.0545(7)  | 0.1322(2)   | 0.8177(4)  | 0.0119(5)  |
| O3   | 0.4351(7)  | 0.1271(2)   | 0.7106(6)  | 0.0172(7)  |
| O4   | 0.0450(7)  | –0.0416(2)  | 0.8131(5)  | 0.0155(6)  |
| O5   | 0.0656(7)  | 0.2958(3)   | 0.7951(6)  | 0.0240(8)  |
| O6   | 0.0572(7)  | –0.2112(2)  | 0.8003(5)  | 0.0218(7)  |

**Table 2.** Interatomic distances (in Å) for Na$_5$Cu$_3$O$_6$.

| Atom  | O1                   | O2       | O3       | O4                   | O5                   | O6                   |
|-------|----------------------|----------|----------|----------------------|----------------------|----------------------|
| Cu1   | 1.936(3)             | 1.936(3) | 1.935(3) | 1.926(3)             |                      |                      |
| Cu2   | 1.903(3)             |          |          | 1.896(4)             | 1.904(4)             | 1.926(4)             |
| Cu3   |                      | 1.881(3) | 1.874(3) |                      | 1.855(4)             | 1.880(4)             |
| Na1   | 2.499(4)             | 2.332(4) | 2.373(4) | 2.335(4) 2.377(4)    |                      |                      |
| Na2   | 2.606(4)             | 2.315(4) |          | 2.374(4)             | 2.316(4)             | 2.340(4)             |
| Na3   |                      | 2.501(5) | 2.703(5) |                      | 2.332(5)             | 2.300(5) 2.318(5)    |
| Na4   | 2.297(5) 2.493(5)    |          | 2.408(5) | 2.299(5)             |                      |                      |
| Na5a  | 2.604(7)             | 2.355(6) | 2.333(6) |                      | 2.270(7)             |                      |
| Na5b  |                      | 2.295(6) | 2.290(6) |                      | 2.397(8) 2.767(8)    | 2.614(7)             |

**Table 3.** Selected bond angles (°) in $Na_5Cu_3O_6$

| atoms | angles (°) | atoms | angles (°) |
|---|---|---|---|
| O(3)-Cu(1)-O(1) | 91.92(15) | O(5)-Cu(3)-O(6) | 84.81(18) |
| O(3)-Cu(1)-O(2) | 84.09(15) | O(3)-Cu(3)-O(6) | 92.83(17) |
| O(4)-Cu(1)-O(1) | 85.67(15) | O(5)-Cu(3)-O(2) | 95.06(16) |
| O(4)-Cu(1)-O(2) | 98.28(15) | O(3)-Cu(3)-O(2) | 87.31(15) |
| O(4)-Cu(1)-O(3) | 177.52(15) | O(5)-Cu(3)-O(3) | 177.26(16) |
| O(1)-Cu(1)-O(2) | 175.11(14) | O(6)-Cu(3)-O(2) | 179.26(17) |
| O(4)-Cu(2)-O(1) | 87.43(15) | Cu(2)-O(1)-Cu(1) | 93.13(16) |
| O(1)-Cu(2)-O(5) | 93.96(17) | Cu(3)-O(2)-Cu(1) | 94.13(15) |
| O(4)-Cu(2)-O(6) | 96.44(16) | Cu(3)-O(3)-Cu(1) | 94.42(16) |
| O(5)-Cu(2)-O(6) | 82.24(17) | Cu(2)-O(4)-Cu(1) | 93.66(17) |
| O(4)-Cu(2)-O(5) | 177.64(18) | Cu(3)-O(5)-Cu(2) | 97.22(18) |
| O(1)-Cu(2)-O(6) | 175.54(15) | Cu(3)-O(6)-Cu(2) | 95.68(17) |
| Cu(1)-Cu(3)-Cu(2) | 179.15(3) | | |
| Cu(2)-Cu(1)-Cu(3) | 176.29(3) | | |
| Cu(1)-Cu(2)-Cu(3) | 178.42(3) | | |

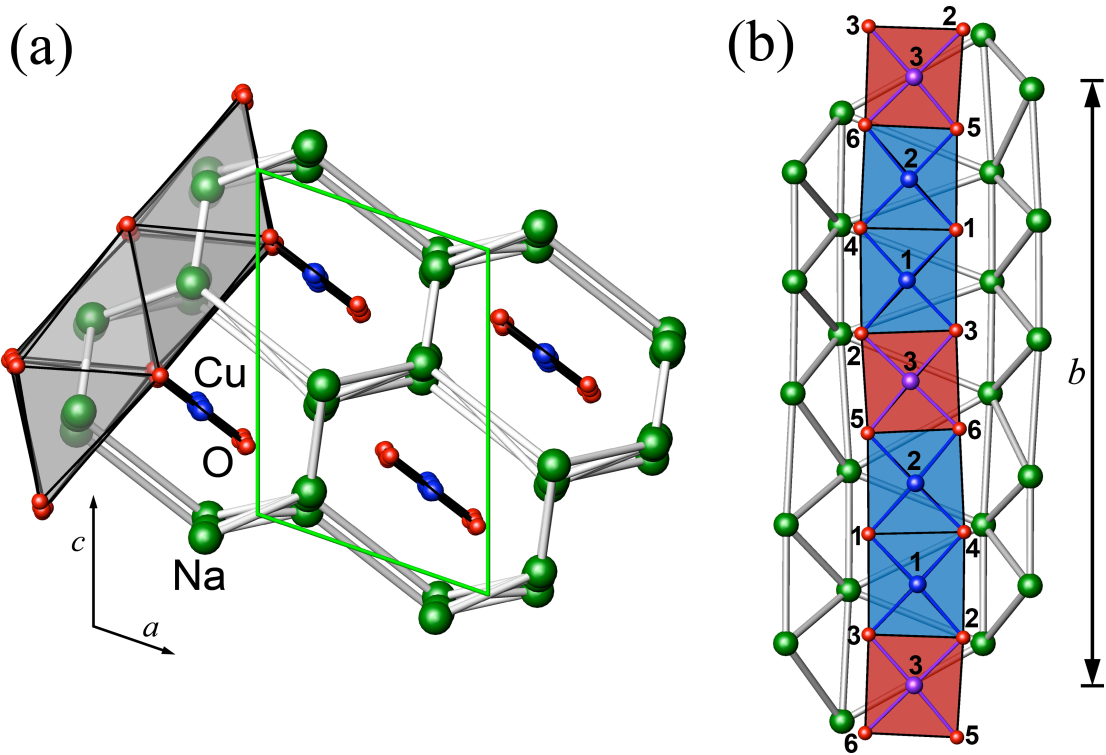

**Fig. 1.** Crystal structure with unit cell (green sticks) of $Na_5Cu_3O_6$: (a) showing the periodicities of Na and $CuO_2$ units with Na ions forming a honeycomb pattern (emphasized by white sticks) with the channels occupied by cuprate ribbons; (b) $CuO_2$ chains in $Na_5Cu_3O_6$: showing the periodicities of $Cu^{3+}$ (red squares) and $Cu^{2+}$ (blue squares) within each ribbon. The copper and oxygen atoms are numbered corresponding to Table 1.

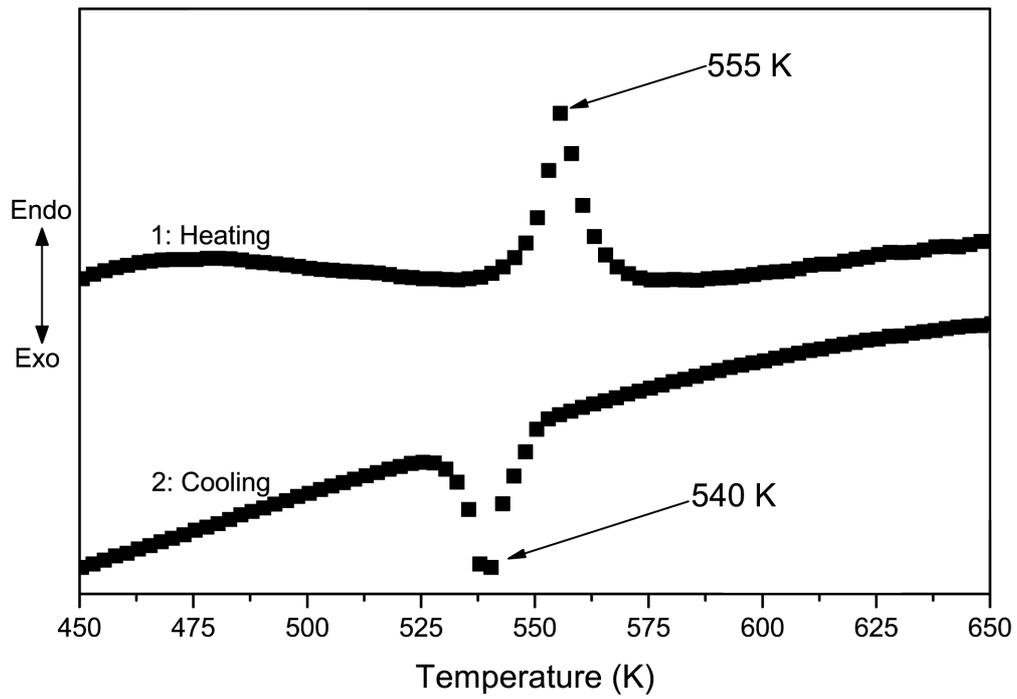

**Fig. 2.** Wigner lattice melting of $Na_5Cu_3O_6$ as detected by differential scanning calorimetry (DSC) at $T$ = 555 (540) K on heating (cooling) respectively.

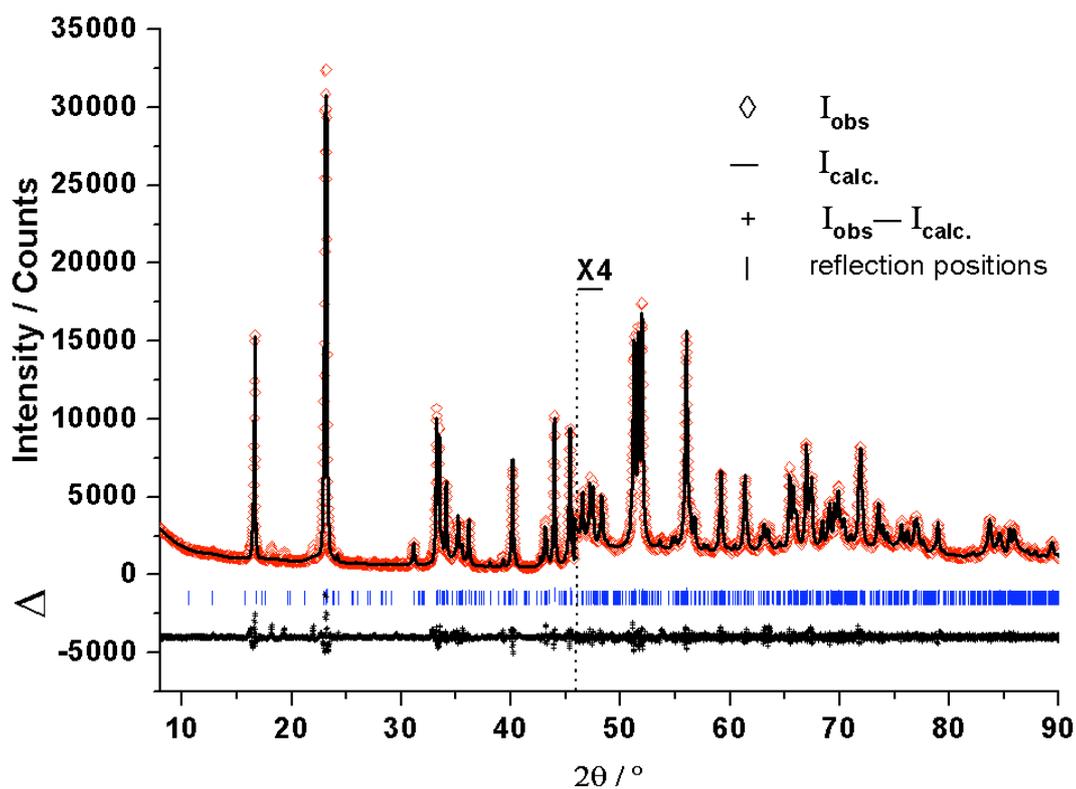

**Fig. 3.** Scattered X-ray intensity for polycrystalline sample of Na$_5$Cu$_3$O$_6$ at $T$ = 298 K as a function of diffraction angle $2\theta$ ($\lambda$ = 1.54059 Å), showing the observed pattern (diamonds), the best Rietveld-fit profile (——) based on single crystal data, reflection markers (vertical bars), and difference plot $\Delta$ = I$_{obs}$-I$_{calc}$ (+) (shifted by a constant amount). Note that the higher angle part is enlarged by a factor of 4 starting at $2\theta$ = 46°.

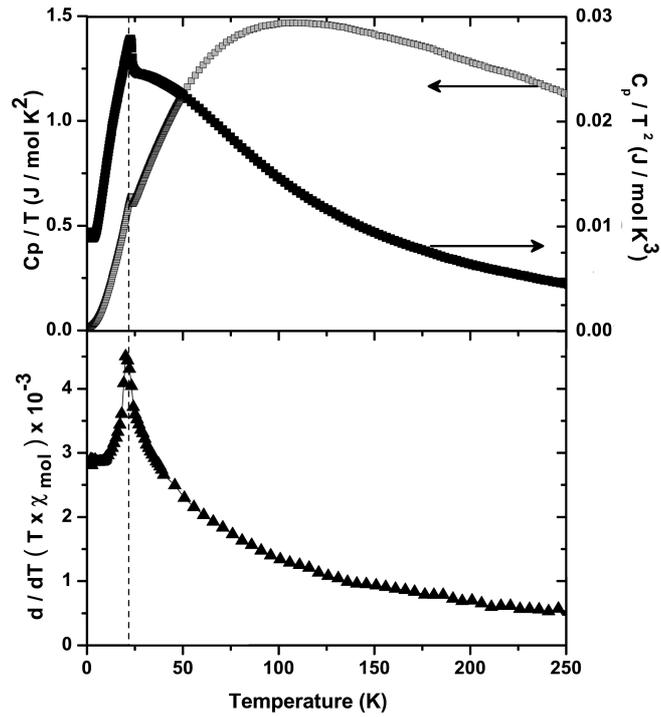

**Fig. 4.** (a) Specific heat ($C_p$/T) at zero field as a function of $T$ of polycrystalline sample of $Na_5Cu_3O_6$. The $C_p/T^2$ plot as a function of $T$ around $T_N$ (panel (a)), and $d/dT$ ($\chi_{mol} \times T$) (Fisher's heat capacity, panel (b)) emphasize the pristinely perfect sample devoid of magnetic defects.

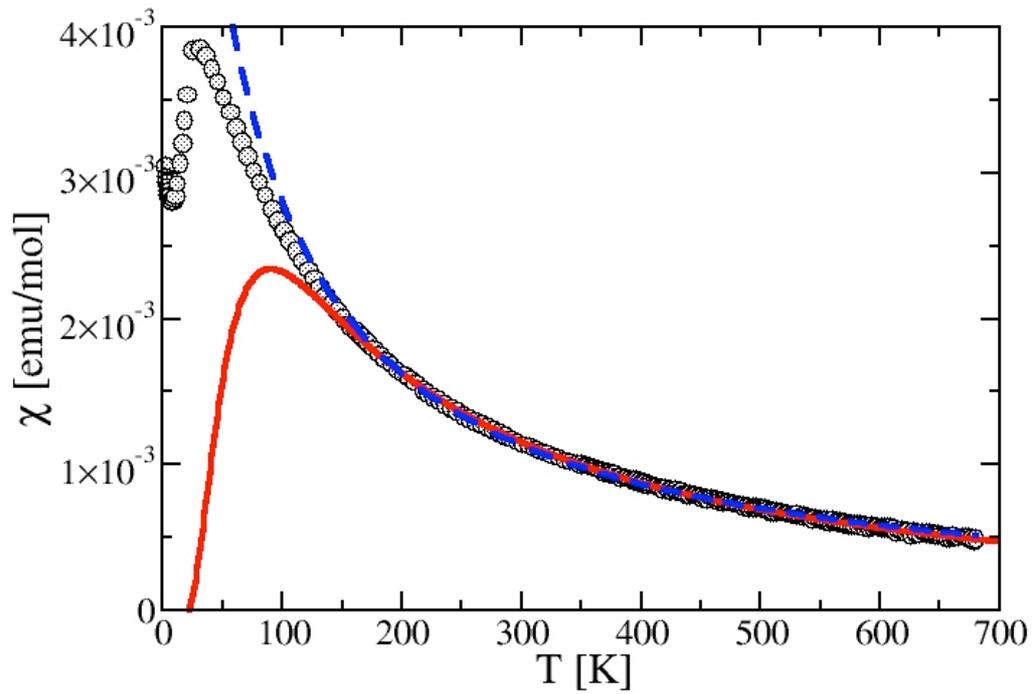

**Fig. 5.** Measured susceptibility (symbols) compared to a Curie law fit (dashed line) for 150 K < $T$ < 680 K with $C$ = 0.40 K emu/mol and $\theta$ = –40.5 K and to an independent dimer model (solid line) with exchange constant $J_2$ = 145 K and $\chi_0$ = –0.9×10$^{-4}$ emu/mol.

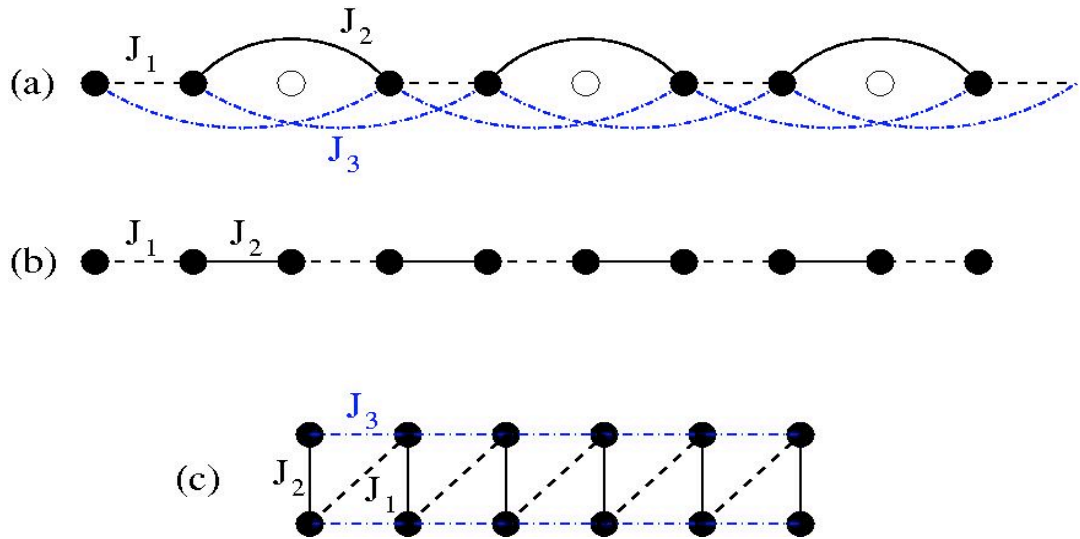

**Fig. 6.** (a) Chain with $Cu^{2+}$ (filled circles, spin ½) and $Cu^{3+}$ ions (open circles, no spins) and magnetic exchange couplings $J_1$, $J_2$, $J_3$. (b) Simplified magnetic structure showing only magnetic $Cu^{2+}$ ions and couplings $J_1$, $J_2$. In this approximation the model is an alternating ferro-antiferromagnetic Heisenberg chain. (c) Taking also $J_3$ into account the magnetic model is equivalent to a Heisenberg ladder.

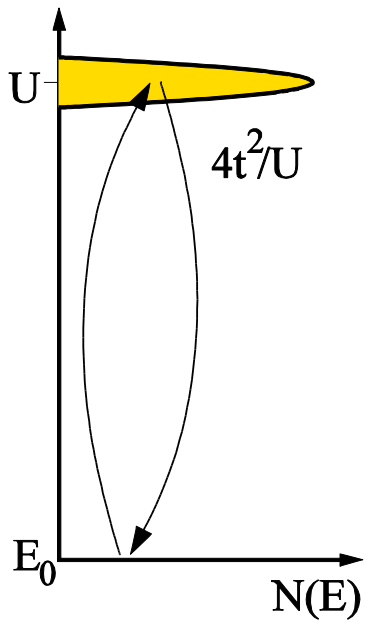 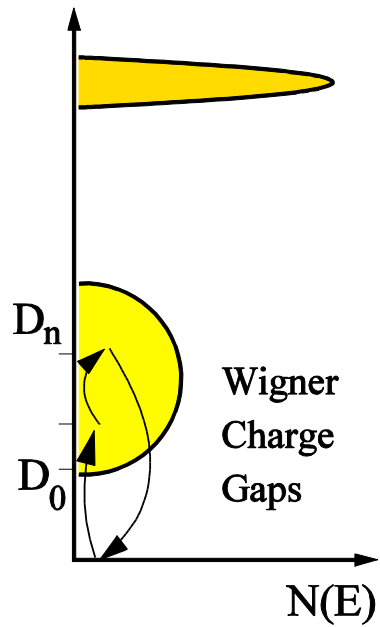 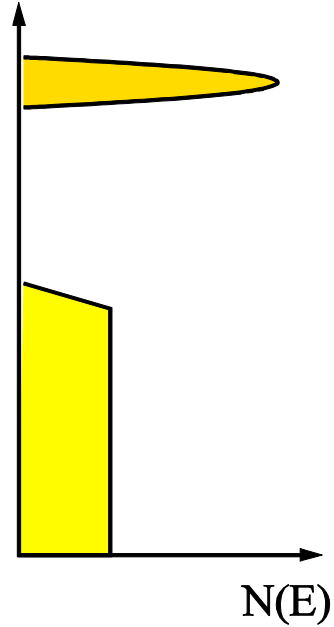

**Fig. 7.** (a) Schematic view of a Mott-Hubbard insulator: Excitations from the ground state at $E_0$ to double occupied configurations in the upper Hubbard band with energy $U$ lead to conventional antiferromagnetic superexchange interaction $\sim 4t^2/U$. (b) The charge excitations $D_n$ of a Wigner crystal are in general small compared to $U$. Higher order processes of such charge excitations can contribute to superexchange with antiferro- or ferromagnetic interactions and thereby compete with the conventional superexchange. (c) When the Wigner lattice melts, e.g. at high temperature, charge gaps disappear and the system changes into a correlated metal.

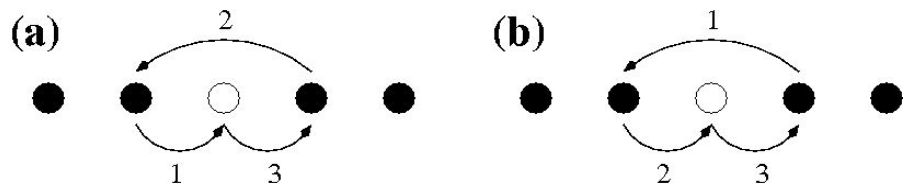

**Fig. 8.** Sequence of electron transitions, 1-2-3, in two distinct exchange processes: (a) Virtual excitation across the Wigner gap giving a contribution $\sim t_1^2 t_2/D^2$ to $J_2$. (b) Superexchange process yielding a contribution $J_2 \sim t_1^2 t_2/(UD)$. Exchange process (b) is a virtual excitation across the Mott-Hubbard gap U while exchange process (a) only involves the Wigner gap D.

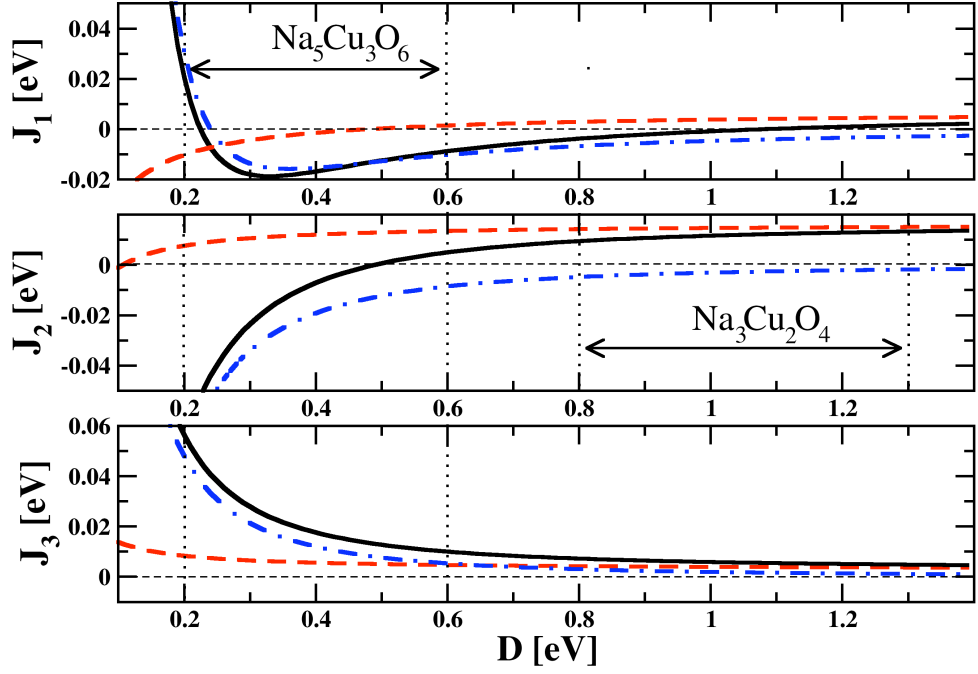

**Fig. 9.** Exchange constants, Eq. (5), as a function of the energy scale for charge fluctuations D for U=3.5 eV, $t_1$ =-0.08 eV, $t_2$ =-0.12 eV, and $t_3$ =0.05 eV. Shown are the total exchange constants (solid lines), the superexchange contributions involving U (dashed lines), and the terms involving virtual excitations across the Wigner gap (dot-dashed lines).

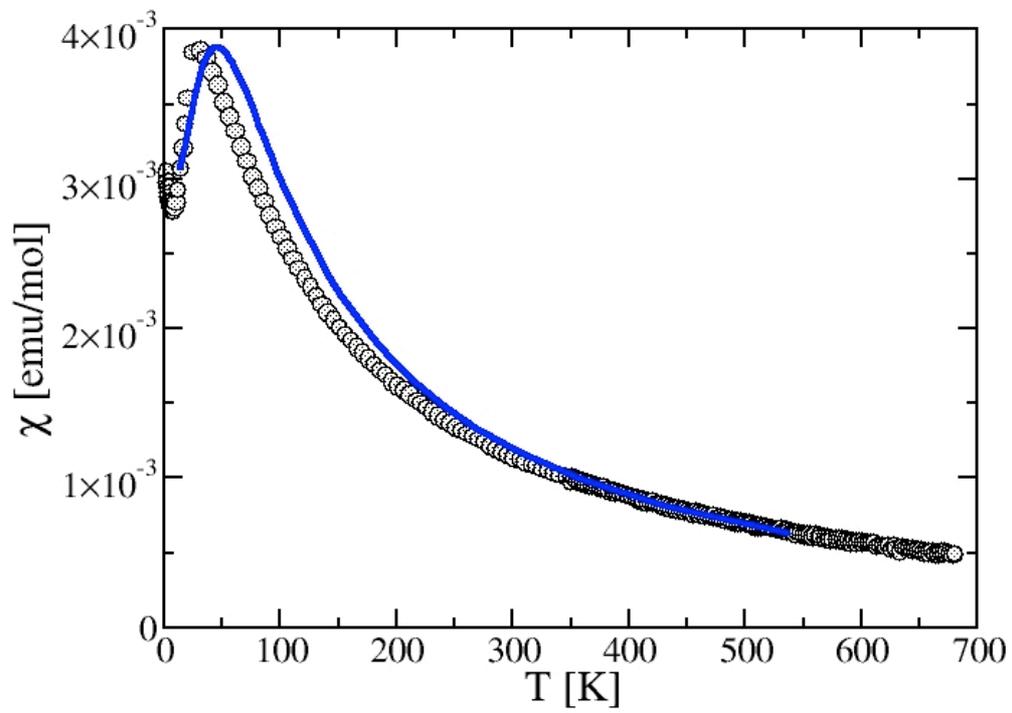

**Fig. 10.** Experimental data for the susceptibility (symbols) compared to the susceptibility for a Heisenberg model with $J_1 = -43$ K, $J_2 = 0$ K, $J_3 = 69$ K, and $\chi_0 = -1.6\times10^{-4}$ emu/mol as calculated numerically using the TMRG algorithm.